\documentclass[11pt]{article}

\usepackage{amsmath}
\usepackage{amssymb}
\usepackage{epsfig}
\usepackage{graphicx}
\usepackage{psfrag}

\usepackage[left=2cm,right=2cm,top=2.5cm,bottom=2.5cm]{geometry}

\def\be{\begin{equation}}
\def\ee{\end{equation}}

\begin{document}

\begin{flushright}
{\large ULB-TH/07-04}
\end{flushright}

\vspace{0.5cm}

\begin{center}

{\Large {\bf  Bound on the Dark Matter Density in the Solar System from Planetary Motions}}
\vspace{0.3cm}\\
{\large  J.-M. Fr\`ere, F.-S. Ling, G. Vertongen}
\vspace{0.3cm}\\

Service de Physique Th\'eorique\\
Universit\'e Libre de Bruxelles\\ B-1050 Brussels, Belgium
\vspace{0.2cm}\\

\end{center}


\abstract{High precision planet orbital data extracted from direct observation, 
spacecraft explorations and laser ranging techniques enable to put a
strong constraint on the maximal dark matter density of a spherical halo centered around the
Sun. The maximal density at Earth's location is of the order $10^5$~${\rm
GeV/cm^3}$ and shows only a mild dependence on the slope of the halo profile, taken between 0 and -2. 
This bound is somewhat better than that obtained from the perihelion precession limits. }

\section*{Introduction}

The existence of Dark Matter in the Universe has been inferred from the observation of
structures much larger than the Solar System. Since Fritz Zwicky first suggested in 1933 the
presence of a non-luminous matter component in the Coma galaxy cluster~\cite{Zwicky:1933gu},
numerous experimental observations as well as theoretical studies have fortified this
paradigm. On the cosmological scale, the analysis of the cosmic microwave background mapped by
the WMAP satellite~\cite{Spergel:2003cb} has led to the so-called Concordance Model of
cosmology~\cite{Ostriker:1995rn}, in which only around $4\%$ of the total energy density is
made of ordinary matter, while the rest is shared between Dark Matter ($26\%$) and Dark Energy
($70\%$). The need for Dark Matter has also been shown in structure formation scenarios and
simulations. On the galaxy cluster scale, the amount of Dark Matter inferred from galaxy
kinematical observations matches the strong lensing data of the Hubble Space
Telescope~\cite{Hudson:1997bj}. In the next future, strong and weak lensing surveys with SNAP
and LSST should provide us with a detailed sky map of the gravitational shear induced by Dark
Matter~\cite{Marshall:2005nv}. On a galactic scale, numerous spiral galaxies have been studied
and yield flat or even rising rotation curves, that extend well beyond the visible
disk~\cite{Sofue:2000jx}. This is usually interpreted as due to a Dark Matter halo, much
bigger than the disk itself.

If Dark Matter is an essential component of the universe, there is some hope to discover it in
our own galaxy, the Milky Way. As expected, modeling and understanding the detailed structure
of the halo becomes crucial to estimate the intensity and the spectrum of the Dark Matter
signal in direct as well as indirect detection experiments. It is generally accepted that the
local Dark Matter density at Earth's location should be around 0.3 GeV/cm$^3$, although a much
higher value is possible in the presence of clumps, or near a caustic~\cite{Sikivie:1997ng}.
Therefore, one can imagine that the Solar System itself is surrounded by a local subhalo, with
a local density well above the galactic value. Several processes have been invoked to clump
the Dark Matter in the Solar System. For example, Dark Matter particles could become trapped
inside the Sun's gravitational potential through multiple scatterings~\cite{Press:1985ug} or
they could be captured in the Solar System as a result of the gravitational pull from
planets~\cite{Damour:1998rh}. It might even be that the existence of the Solar System itself
is evidence for a local subhalo. In any case, a Dark Matter halo centered around the Sun will
influence the motion of the planets. The purpose of this paper is to put limits on such a
possibility.

It is worth emphasizing that the presence of Dark Matter is NOT needed on the scale of the
Solar System, despite all the aforementioned evidences at larger scales. Celestial Mechanics
is one of the greatest achievement of newtonian mechanics, that enabled to calculate
ephemerides tables with high precision, predict eclipses, and even led to the discovery of
Neptune in 1846 and of Pluto in 1930. Of course, we now know that newtonian mechanics is not
the whole story, as people looked for a hypothetical planet to explain Mercury's perihelion
precession, but never found it...

Nowadays, planet orbits have been so precisely measured~\cite{Pitjeva:2005} that they are able to constrain the
local Dark Matter density. Previous estimates give bounds of the order of
10$^{-19}$ to 10$^{-20}$ g/cm$^3$, and are derived from observational limits on the secular
perihelion precession of the inner planets ({\it i.e.} the residual precession of a planet
after all known effects, like the general relativity contribution, or the precession due to
the influence of other planets are
subtracted)~(\cite{Khriplovich:2006sq},\cite{Iorio:2006cn},\cite{Sereno:2006mw}). In this
paper, we point out that the presence of a Dark Matter halo centered around the Sun will not
only cause the planets to precess, but will also alter all the keplerian parameters used to
fit a planet's orbit. 
In particular, if the halo density is high enough, the third Kepler's law cannot hold for all planets,
even when all the uncertainties of each planet's orbital parameters are taken into account.  
Current planetary orbit determinations are precise enough to provide a meaningful constraint.
Of course, at such precision, a planet orbit cannot be viewed as a pure keplerian elliptic orbit anymore,
as the mutual influence of planets introduces many periodic or secular corrections to the keplerian parameters. 
For our purpose, it is however enlightening to continue to discuss the problem in terms of
keplerian parameters, in the spirit of a perturbation expansion, where the different perturbations 
are added separately.    

Hereafter, we will therefore describe the orbit of a planet, which is approximately elliptic, by
only three parameters, which are the semi-major axis $a$, the eccentricity $e$, and the revolution period $T$.
In the presence of a Dark Matter halo, we first notice that the third Kepler's law can always hold for one planet, 
irrespective of the density of the halo (but still small enough compared to the mass of the Sun
so that perturbation scheme is valid), because the orbital motion can only probe the total mass interior to the orbit.
However, with several planets, the halo density is directly limited by the precision at which the third Kepler's law
is verified.  

The paper is organized as follows. In section~1, we summarize the salient features of the
keplerian orbit. We recall the various relations between the dynamical variables and the
measurable parameters. In section~2, we show how the dynamical variables get modified by the
introduction of a radial perturbation. In particular, we calculate the shift in the
gravitational constant $\Delta K$ and the shift of the perihelion $\Delta \Theta$. Our method also provides a useful 
formula for the perihelion precession. In section~3, these analytical results are
applied to a heliocentric spherical Dark Matter halo and numerical estimates are derived and discussed.

\section{The keplerian orbit revisited}

The relative motion of a planet of mass $m_P$ around the Sun (with mass $M_\odot$) is usefully
parameterized by 6 parameters: $K_0 \equiv G(M_\odot + m_P)$ gives the strength of the
keplerian potential $U_0(r) = -K_0/r$, $L_0$ is the angular momentum per unit mass, $E_0$ is
the energy per unit mass (in absolute value), the semi-major axis $a$ and the eccentricity $e$
characterize the elliptical orbit, and $T$ is the revolution period around this orbit. All of
them are not independent, as
\begin{eqnarray}
K_0 &=& 4 \pi^2 \frac{a^3}{T^2}~,
\label{kepler3}\\
E_0 &=& \frac{K_0}{2a}~,\\
L_0^2 &=& K_0 p~,
\end{eqnarray}
where $p=a(1-e^2)$ is called the semilatus rectum. The subscript $0$ under a parameter is used
to designate its value in the unperturbed scheme ({\it i.e.} without halo). For given values
of $K_0$, $L_0$, and $E_0$, the orbital elements $a$, $e$ and the period $T$ can be
calculated. The period $T$ is found by computing \be T = 2 \int_{r_{min}}^{r_{max}}
\frac{dr}{\sqrt{-2\left( E_0 + U_0(r) \right) - L_0^2/r^2}}~, \ee where $r_{min}=a(1-e)$ and
$r_{max}=a(1+e)$ are respectively the orbit's perihelion and aphelion distances to the Sun. In
this integral, $r_{min}$ and $r_{max}$ are calculated as the roots of the function \be
f_0(r)=-2\left( E_0 + U_0(r) \right) - \frac{L_0^2}{r^2}~. \ee The third Kepler's law
(eq.~(\ref{kepler3})) is obtained by using the following parameterization of the elliptical
orbit \be r = a (1-e \cos \xi)~, \ee where $\xi$ is called the eccentric anomaly parameter.

To calculate the angle variation around the orbit, \be \Theta_0 = 2\int_{r_{min}}^{r_{max}}
\frac{L_0/r^2~dr}{\sqrt{-2\left( E_0 + U_0(r) \right) - L_0^2/r^2}}~, \ee the parameterization
\be r  = \frac{p}{1 + e \cos \theta}~, \ee where $\theta$ is the polar angle, also called the
true anomaly parameter, proves to be more convenient. For a pure keplerian motion, $\Theta_0$
is of course $2\pi$, as there is no perihelion precession.

The planets parameters are summarized in table~\ref{parameters1}.

\section{Shift of the keplerian parameters due to a  perturbation in the potential}

In this section, we will derive how the keplerian parameters are affected by a perturbation in the potential,
\be
U(r) = U_0(r)+U_1(r)~,
\ee
which is supposed to be small, $U_1(r) \ll U_0(r)$ along the unperturbed orbit.

As discussed in the introduction, the orbital elements $a$, $e$ and the period $T$ of the
planets in the Solar System are known to a high precision. We will estimate in section~3 the
uncertainty on our results allowed by a small variation of these parameters. In this section,
however, we will keep $a$, $e$ and $T$ fixed. As a consequence, the dynamical parameters $E$,
$L$ and $K$ will necessarily undergo a shift after the introduction of the perturbation. We
will write
\begin{eqnarray*}
E&=&E_0 + \Delta E~,\\
L^2&=& L_0^2 + \Delta L^2~,\\
K&=& K_0 + \Delta K~,
\end{eqnarray*}
and
\be
f(r) = -2\left( E + U(r) \right) - \frac{L^2}{r^2} = f_0(r) + \Delta f(r) ~.
\ee
Notice that $\Delta E$ and $\Delta L^2$ can be expressed in terms of $U_1(r)$ and $\Delta K$ by requiring
the invariance of the orbital elements. The period is then calculated as a function of $\Delta K$,
\be
T(\Delta K) = 2 \int_{r_{min}}^{r_{max}} \frac{dr}{\sqrt{f_0(r) + \Delta f(r,\Delta K)}}
\simeq T_0 - \int_{r_{min}}^{r_{max}} \frac{\Delta f(r,\Delta K)~dr}{f_0(r)^{3/2}}~,
\ee
where the last integral is converging by construction.
By requiring the invariance of the period, one finds
\be
\Delta K = \frac{3a}{2e} \left[ (1+e)^2 U_1(r_{max}) - (1-e)^2 U_1(r_{min}) \right] +
\frac{2}{\pi a e} \int_0^{\pi} V'_1(r(\xi))\cos \xi ~d\xi~,
\label{Kshift}
\ee
with $V_1(r) = r^3 U_1(r)$, and the derivation is taken with respect to $r$.

Our methodology also enables us to calculate the perihelion shift per revolution, \be \Delta
\Theta \simeq \pi \frac{\Delta L^2}{L_0^2} - \int_{r_{min}}^{r_{max}} \frac{\Delta f(r,\Delta
K)~L_0/r^2~dr}{f_0(r)^{3/2}} = \frac{2}{e K_0}\int_0^{\pi} U_1'(r(\theta)) ~r^2(\theta) \cos
\theta ~d\theta~. \label{perishift} \ee It is worth noticing that the perihelion shift is
independent of $\Delta K$, {\it i.e.} independent of a variation in the keplerian potential,
as expected. Furthermore, eq.~(\ref{perishift}) is equivalent to the Landau \& Lifschitz
formula~\cite{Landau} \be \Delta \Theta = \frac{\partial}{\partial L_0} \left ( \frac{2}{L_0}
\int_0^{\pi} r^2(\theta) U_1(r(\theta)) ~d\theta \right ) ~, \ee although this may not be
obvious at first sight, which we checked numerically.

\section{Constraints on a heliocentric Dark Matter halo}

In this section, we examine the consequences on the planetary motions of the presence of a
spherically symmetric Dark Matter halo centered around the Sun, with a density profile defined
by \be \rho (r) = \rho_0 \left(\frac{r}{r_0}\right)^{-\gamma}~, \ee where we take $0\leq
\gamma <2$, and $r_0$ is the mean Earth to Sun distance, so that $\rho_0$ is the Dark Matter
density at Earth's location. $\gamma =0$ corresponds to a constant density halo. Note that
such halo will affect the orbits of the planets only if it is heliocentric. $\gamma=1$
corresponds to the NFW profile that appears systematically in Dark Matter
simulations~\cite{Navarro:1996gj}.

The halo mass enclosed in a sphere of radius $r$ is equal to
\be
M(r) = \frac{4 \pi \rho_0}{(3-\gamma)r_0^{-\gamma}}~r^{3-\gamma}~.
\ee
It will correspond to a gravitational potential
\be
U_1(r) = \frac{4 \pi G \rho_0}{(3-\gamma)(2-\gamma)r_0^{-\gamma}} ~r^{2-\gamma}~.
\ee
The shifts of $\Delta K$ and the perihelion given by eqs.~(\ref{Kshift})~and~(\ref{perishift}) reduce to
\begin{eqnarray}
\Delta K (\rho_0,\gamma) &=& -(4-\gamma) G M(a)~, \label{gammashift1}\\
\Delta \Theta(\rho_0,\gamma) &=& -\pi(3-\gamma) \frac{M(a)}{M_{\odot}}~.
\label{gammashift2}
\end{eqnarray}
Notice that eq.(19) differs from the results for $\gamma=0$ in 
\cite{Khriplovich:2006sq},\cite{Iorio:2006cn}, but agrees with~\cite{Sereno:2006mw}.

These shifts provide a constraint on the Dark Matter density at Earth's location
when compared to the experimental limits found in table~\ref{parameters2}.
Notice that the values listed for $\Delta \Theta$ and $\Delta a$ are the result 
of a general $\chi^2$ orbit fitting of a huge amount of 
astronomical data, and therefore cannot be considered as independent from each other.
It is curious that the residual perihelion shift for Venus is not compatible with zero.
Also, there is no data on Neptune's perihelion precession
due to the fact that Neptune's revolution period is quite long compared to the time span its
orbit has been observed, {\it i.e.} 90 years.
The rationale for deriving a constraint from the calculated values of $\Delta K$ goes as follows.
Essentially, the maximal shift is limited by the precision at which the period and the semi-major axis 
of a planet are known, by application of the third Kepler's law. Now, the relative precision of these
two elements are reasonably of the same order of magnitude, as they arise from the same fitting procedure.
Also notice that the masses of the planets are determined with great accuracy by spacecraft probes~\cite{Standish:1995}.
Therefore, the maximal shift is limited by the orbit precision combined with a possible modification of the solar mass.
As already mentioned in the introduction, the knowledge of the orbital elements $a$ and $T$ for one
planet is not sufficient to constraint the halo density, as the shift $\Delta K$ can always be absorbed
in a modification of $M_\odot$. For several planets, however, the difference of the measured value of $K$ is a direct
probe of the halo mass. 

\begin{table}[!htb]
\caption{Planets keplerian parameters $a$, $e$, $T$ and $K_P \equiv G m_P$ (from \cite{Standish:1995}).
Notice that $G M_\odot=1.327~10^{20}~m^3s^{-2}$.}\label{parameters1}
\begin{center}
\begin{tabular}{|c|c|c|c|c|}\hline
Planet & Period (days) & $a$ (AU) & $e$ & $K_P$ ($m^3 s^{-2}$)\\
\hline \hline Mercury &87.96935&0.38709893&0.20563069 & $2.20~10^{13}$\\
\hline Venus&224.70096&0.72333199&0.00677323 & $3.25~10^{14}$\\
\hline Earth&365.25696&1.0000001124&0.01671022 & $4.04~10^{14}$\\
\hline Mars&686.9601&1.52366231&0.09341233 & $4.28~10^{13}$\\
\hline Jupiter&4335.3545&5.20336301&0.04839266 & $1.27~10^{17}$\\
\hline Saturn&10757.7365&9.53707032&0.05415060 & $3.79~10^{16}$\\
\hline Uranus&30708.16002&19.19126393&0.04716771 & $5.79~10^{15}$\\
\hline Neptune&60224.9036&30.06896348&0.00858587 & $6.84~10^{15}$\\
\hline
\end{tabular}
\end{center}
\end{table}

\begin{table}[!htb]
\caption{Experimental limits (from \cite{Pitjeva:2005},\cite{Pitjeva:2005b},\cite{Standish:1995})}\label{parameters2}
\begin{center}
\begin{tabular}{|c|c|c|c|c|}\hline
Planet & $\Delta \Theta$ (arcsec/century) & $\Delta a$ (m) & $\Delta K_P$ ($m^3 s^{-2}$)\\
\hline \hline Mercury &-0.0036(50)& 0.105 & $9.14~10^{8}$\\
\hline Venus&0.53(30)& 0.329 & $6.0~10^{6}$\\
\hline Earth&-0.0002(4)& 0.146 & $2.45~10^{7}$\\
\hline Mars&0.0001(5)& 0.657 & $2.8~10^{5}$\\
\hline Jupiter&0.0062(360)& 639 & $9.68~10^{8}$\\
\hline Saturn&-0.92(2.9)& 4222 & $1.08~10^{9}$\\
\hline Uranus&0.57(13.)& 38484 & $7.59~10^{9}$\\
\hline Neptune&no data& 3463309 & $1.4~10^{10}$\\
\hline
\end{tabular}
\end{center}
\end{table}

In figure~1, the maximal allowed Dark Matter density at Earth's location from the $K$ shift
constraint and from the perihelion bounds is plotted against $\gamma$.
\begin{figure}[!htb]
\begin{center}
\psfrag{g}[c][c]{\tiny $\gamma$}
\psfrag{rho}[c][c]{\tiny $\log_{10}\rho_0^{max}~({\rm GeV/cm^3})$}
\psfrag{peri}[c][c]{\tiny $\Delta \Theta$ constraint}
\psfrag{K}[c][c]{\tiny $\Delta K$ constraint}
\includegraphics[height=.45\textwidth]{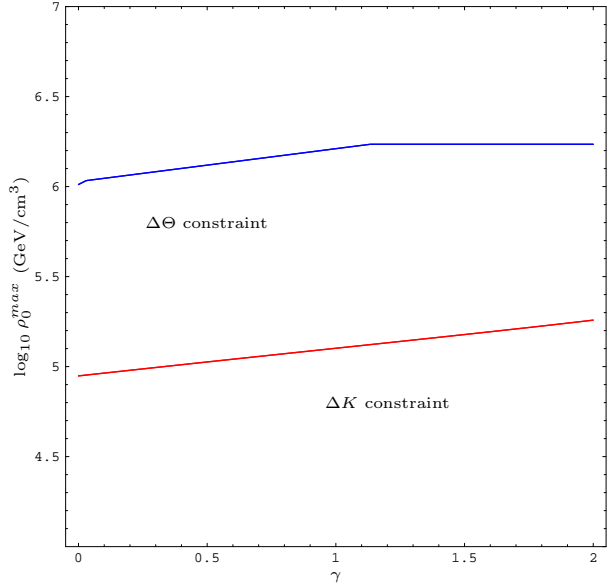}
\caption{Maximal allowed Dark Matter density at Earth's location, as a function of the halo
profile parameter $\gamma$}
\end{center}
\end{figure}
We can see that the first constraint is somewhat more stringent. 
We also notice that the maximal density does not change much when $\gamma$ is varied between 0 and 2,
because it is determined by the precision of the Earth's and Mars' orbits, which are the most constraining. 

Finally, let us discuss the so-called Pioneer anomaly in the context of a heliocentric Dark Matter halo.
It consists of an unexplained constant and uniform acceleration
\begin{eqnarray}
a_P = (8.74 \pm 1.33) \cdot 10^{-10} ~\mbox{m/s$^2$}.
\end{eqnarray}
directed towards the Sun and detected by spacecrafts Pioneer 10 and 11 
at a distance between 20 AU and respectively 70 and 40 AU~\cite{Anderson:1998jd} from the Sun. 
Such an acceleration would correspond to a halo profile characterized by $\gamma=1$ and $\rho_0 \approx 8\cdot 10^9$~GeV/cm$^3$,
which would have dramatically altered the planetary orbits in the solar system. 
The Pioneer anomaly cannot therefore be explained with a spherical halo with monotonically decreasing density.

\section*{Conclusion}

We have evaluated the bound on the maximal allowed Dark Matter density at Earth's location from
planetary data using the perihelion and the $K$ shifts. 
The latter bound is somewhat better, although such precision has to be considered as quite optimistic.
Conservative bounds could be one order of magnitude higher.
We have also derived a useful expression for the computation of the perihelion shift.

\section*{Acknowledgments}

We would like to warmly thank E.M. Standish, P. Rosenblatt and N. Rambaux for helpful discussions.
This work is supported by the FNRS, I.I.S.N. and the Belgian Federal Science Policy (IAP 5/27).

\end{document}